Spin Rotation by Earth's Gravitational Field in a "Frozen-Spin" Ring

Yuri Orlov [a,1], Eanna Flanagan [b], and Yannis Semertzidis [c]

[a] *Department of Physics, Cornell University, Ithaca, NY 14853, USA*

[b] *Departments of Physics and Astronomy, Cornell University, Ithaca, NY 14853, USA*

[c] *Brookhaven National Laboratory, Upton, NY 11973-5000, USA*

[1] Corresponding author: Dept. of Physics, 310 Physical Sciences Building, Cornell University, Ithaca, NY 14853, USA. Tel: +1 607 255 3502; fax: +1 607 254 4552; e-mail: yfo1@cornell.edu.

**Abstract**

Detailed calculations of spin rotation by the Earth's gravitational field in a frozen-spin ring are presented in three different coordinate systems and used (a) to show that the systematic error caused by gravitation in a proposed electric dipole moment measurement can be unambiguously determined, and (b) to propose measuring the spin-gravity effect in a dedicated frozen-spin ring using electrons.

# 1. Introduction and basic equations

This paper was chiefly motivated by proposals to measure the electric dipole moments (EDMs) of various charged particles in "frozen-spin" storage rings [1], [2], [3]. EDM here means a vector, $\vec{d}$, whose rest-frame energy in an electric field equals $U = -\vec{d} \cdot \vec{E}$ (as it does for any classical electric dipole), but whose direction is defined by a particle's rest-frame spin, $\vec{s}$, $\vec{d} = (e\hbar/2mc)\eta\vec{s}$. Such a $\vec{d}$ violates T- and P-symmetries and has never been directly observed. The Standard Model predicts EDMs at a level much below $10^{-29} e \cdot cm$, the current hoped-for sensitivity of proposed proton and deuteron EDM measurements. However, SUSY and some other physical models predict much larger ones. Discovery of an EDM at the level of $d > 10^{-29} e \cdot cm$, which for protons corresponds to $\eta_p > 2 \times 10^{-15}$, would obviously open a window to new physics. At that level of accuracy, unavoidable perturbations like spin rotation by the Earth's gravitational field and by the Earth's rotation must be seriously investigated as sources of systematic error in EDM measurement. Our paper focuses on the leading gravitation spin effect because, in a ring, the Earth's rotation is not nearly as significant a source of error as the Earth's gravitational field. The effect of the Coriolis force, for example, averages to zero due to particle revolutions. (See Section 5 for further discussion of this point.)

A frozen-spin ring is designed so that the initial polarization of a particle relative to the equilibrium orbit will remain constant ("frozen") in the absence of EDM. Then, ideally, only the EDM will rotate the spin—in the plane perpendicular and tangential to that orbit. This requires a combination of particle momentum $p$, $p/mc = \beta\gamma$, $\gamma = \varepsilon/mc^2$, and labframe vertical magnetic $B$-field and radial electric $E$-field such that,

$$\vec{\omega}_a = -\frac{e}{mc}\left[a\vec{B} + \left(a - \frac{1}{\gamma^2\beta^2}\right)\frac{\vec{E} \times \vec{v}}{c^2}\right] = 0. \tag{1.1}$$

$\omega_a$ is the rotation frequency of the *planar* spin projection relative to the rotation frequency of the momentum [1]; $a = (g-2)/2$ where $g$ is the gyromagnetic factor. When $\omega_a = 0$, the planar angle between $\vec{p}$ and $\vec{s}$ remains constant in time. Some feedback system is needed to hold this condition experimentally. We assume the existence of such a system, as well as a system canceling spin rotations in the vertical plane by perturbations other than EDM. (We refer the reader to [4] for discussion of these non-EDM problems, including betatron and synchrotron stability issues.)

In a frozen-spin EDM ring, the EDM detector(s) measure the vertical spin component as a function of time,

$$\frac{d\vec{s}}{dt} = \frac{\eta e}{2mc}\left(\vec{E} + \vec{\beta} \times \vec{B}\right) \times \vec{s}. \tag{1.2}$$

The initial spin polarization should be along the orbit. If (1.1) is satisfied, the longitudinal spin component, $s_l$, on the right side of (1.2) as well as $E_r$, $\beta_l$, and $B_z$ remain constant, so $s_z$ is growing.

Gravitation is involved here because, among other things, it rotates the spin in the same vertical plane tangential to the orbit. Rough preliminary estimates show that for the case of protons or deuterons, the scale of the EDM spin rotation in the vertical plane at $d = 10^{-29} e \cdot cm$ can be $\sim 10$ nrad/sec, depending on the accessible electric field. The scale of the spin rotation by the Earth's gravitational field in the same plane is $g/c = 30$ nrad/s, which corresponds to a large systematic error in the EDM measurement.

As follows from (1.1), it is possible to use a purely electric frozen-spin ring for particles with a positive anomalous magnetic moment, $a = (g-2)/2 > 0$, which includes protons, muons and electrons but not deuterons, whose $a < 0$. In this case, $B = B_z = 0$ in eq. (1.1) and, obviously, the particle momentum must equal $p = p_{mag} = mc/\sqrt{a}$ to freeze the spin. This is the so-called magic momentum introduced in 1970 for the muon g-2 experiment [5]. For protons, $p_{mag} = 0.7007 \, GeV/c$; for muons, $p_{mag} = 3.09 \, GeV/c$; for electrons, $p_{mag} = 15.005 \, MeV/c$. Such a regime has the following big advantage. Two beams of particles can be injected and rotated in opposite directions, meeting the same perturbations simultaneously, and the corresponding systematic errors can be reduced by combining the clockwise and counterclockwise data [2]. This idea is based on the fact that most perturbations (including the gravitational field) do not violate T-symmetry, whereas the EDM does. In the deuteron case, the magnetic field cannot be zero and the momentum cannot be magic. One can nevertheless try to extract the true EDM value by using clockwise and counterclockwise beams *in turn*, changing the particles' initial conditions and the sign of the magnetic field accordingly [3].

In Sections 2-4 we calculate gravitational spin rotations for the general case of a frozen-spin ring, including the purely electric version, investigating the same spin-gravity effect in different coordinate systems. This approach addresses a problem apparently unremarked in the spin-gravitation literature concerned with rings (see [6], for example): that the measured magnitude of spin rotation in the vertical plane is very sensitive to one's definition of "vertical" *direction*, which itself is determined by the experimental setup. Thus, computations of the spin-gravity effect in different coordinate systems can (as they do here) yield different results, because the vertical directions of the different coordinate systems do not coincide. We need to establish the scale of this unwelcome sensitivity, its sources, and ways to avoid it. We also need to address a parallel question about sensitivity to the choice of "vertical" *position* when the "vertical" direction is given.

Our calculation technique is based on manifestly covariant relativistic spin equations in the framework of Riemannian geometry, with the Schwartzschild metric linearized with respect to $g$, $g = 9.80665 m/s^2$, and small deviations from the circular equilibrium orbit. (Similar covariant equations were used much earlier in [7] with $g = 0$, but in connection with spin.) These equations are the general relativity extension of the quasi-classical Thomas-BMT equation [8] into which we introduce the EDM term:

$$\frac{DS^i}{cD\tau} \equiv \frac{dS^i}{cd\tau} + \Gamma^i_{kl} S^k u^l = \frac{e(1+a)}{2mc^2}\left(F^i_{\ k} S^k + u^i F^{kl} S_k u_l\right) - u^i S_k \frac{Du^k}{cD\tau} + \frac{e\eta}{2mc^2}\varepsilon^{iklm} F_{kj} S_l u^j u_m, \quad (1.3)$$

together with the Lorentz equation,

$$\frac{Du^i}{Dc\tau} \equiv \frac{du^i}{dc\tau} + \Gamma^i_{kl} u^k u^l = \frac{e}{mc^2} F^i_{\ k} u^k, \quad (1.4)$$

where

$$\Gamma^i_{kl} = \frac{1}{2} g^{im}\left(\frac{\partial g_{mk}}{\partial x^l} + \frac{\partial g_{ml}}{\partial x^k} - \frac{\partial g_{kl}}{\partial x^m}\right), \quad (1.5)$$

$\varepsilon^{iklm} \equiv e^{iklm}/\sqrt{-|g|}$; $|g|$ is the determinant of $g_{ik}$. We adopt the convention that Roman indices run over 0, 1, 2, 3. The spacetime interval is $ds \equiv cd\tau$, $ds^2 = g_{ik} dx^i dx^k$. For the purpose of this paper, we need to know only $\Gamma^2_{00}$, $\Gamma^2_{11}$, $\Gamma^3_{00}$, $\Gamma^3_{11}$ in order to calculate the spin rotation by gravitation in the tangential vertical planes of a frozen-spin ring. The components $\Gamma^k_{i2}$, $\Gamma^k_{i3}$ are not needed because, in the equations of motion we use, they are multiplied by $u^2$, $S^2$, $u^3$, $S^3$, which by design equal either zero or almost zero in the equilibrium. $\Gamma^k_{01} = 0$ because our $g_{ik}$ do not depend on $x^0$, $x^1$ and $g_{01} = 0$. $\Gamma^0_{00} = \Gamma^0_{11} = \Gamma^1_{00} = \Gamma^1_{11} = 0$, in our case.

The last term of (1.3) represents the EDM. Although the main actor in the EDM phenomenon is the dual EM field tensor $e^{iklm}F_{lm}/2$, additional factors $u^j$ and $u_m$ must be included in the last term to satisfy conditions $u_i S^i = 0$ and $S_i S^i = constant$. The next-to-last term of (1.3) describes the Thomas precession [9]. Combining this term with the second term in parentheses gives us the generalized BMT lab-frame spin equation,

$$\frac{dS^i}{dc\tau} + \Gamma^i_{kl} S^k u^l = \frac{e}{mc^2}\left[(1+a)F^i_k S^k + au^i(F_{lm}S^l u^m)\right] + \text{EDM}; \quad (1.6)$$

$a = (g-2)/2$, where $g$ is the particle gyromagnetic factor. (Below, we will use notation $g$ only for gravity acceleration and $|g|$ for $\det g_{ik}$.)

For the metric $g_{ik}$ we use the Schwarzschild solution in Schwarzschild coordinates [10]:

$$ds^2 = (1 - R_g/R)c^2 dt^2 - R^2(\sin^2\theta d\phi^2 + d\theta^2) - dR^2/(1 - R_g/R), \quad (1.7)$$

where $R_g = 2kM/c^2$ is the Earth's gravitational radius; $-kM/R = \Phi = \Phi_0 + g\bar{x}^3$, $\bar{x}^3 = R - R_0$, see Fig. 1; $\Phi$ is the gravitational potential, $2\Phi/c^2 \ll 1$; and $\Phi = \Phi_0$ is the potential at the Earth's radius, $R_0$:

$$\Phi_0/c^2 = -gR_0/c^2, \quad R_g/R_0 = 2gR_0/c^2. \quad (1.8)$$

$R_0$ is defined by the Earth's circumference, $L = 2\pi R_0$. The equations for our electro-magnetic fields in a curvilinear coordinate system with gravitation are:

$$\frac{\partial}{\partial x^k}\left(\sqrt{-|g|}F^{ik}\right) = 0, \quad (1.9)$$

$$\frac{\partial F_{ik}}{\partial x^l} + \frac{\partial F_{li}}{\partial x^k} + \frac{\partial F_{kl}}{\partial x^i} = 0 \ . \tag{1.10}$$

## 2. Spin rotation: cylindrical coordinates

What we will call cylindrical coordinates can be defined experimentally in two steps: the direction $\bar{x}^3$ (opposite to the direction of $\vec{g}$) is established and then rotated to $x^3$ (in the plane perpendicular to the orbit) by the known angle $\theta_0$, where $\sin\theta_0 = r_0 / R_0$, and $r_0$ is the radius of the ring, see Fig. 1. The coordinates corresponding to these operations, which are slightly different from the usual cylindrical coordinates, $\varphi, r, z$, are:

$$x^1 = r_0 \varphi, \ x^2 = r - r_0, \ x^3 = z - z_0 \ , \tag{2.1}$$

where

$$r = R\sin\theta, \ z = R\cos\theta ; \tag{2.2}$$

$r_0 = R_0 \sin\theta_0$, $z_0 = R_0 \cos\theta_0$. The labframe spin projections used in the eqs. below are $S^1 \approx \gamma s^1$, $S^0 \approx \gamma \beta s^1$ and the 4-velocities $u^1 \equiv dx^1 / cd\tau \approx p/mc$, $u^0 \equiv dt/d\tau = \gamma$. To compute the metric in the coordinates (2.1), we first rewrite (1.7) using the standard cylindrical coordinates $t, \varphi, r, z$, in which

$$R^2 d\theta^2 + \left(1 - R_g/R\right)^{-1} dR^2 = \left(\frac{z}{R}dr - \frac{r}{R}dz\right)^2 + \left(\frac{r}{R}dr + \frac{z}{R}dz\right)^2 \left(1 - \frac{R_g}{R}\right)^{-1} =$$

$$= \frac{\left[r^2 + z^2\left(1 - R_g/R\right)\right]dr^2 + \left[z^2 + r^2\left(1 - R_g/R\right)\right]dz^2 + 2rz\left(R_g/R\right)dr\,dz}{R^2\left(1 - R_g/R\right)} \ . \tag{2.3}$$

Further on, we neglect terms proportional to $g^2$. For example,

$$\left[r^2 + z^2\left(1 - R_g/R\right)\right]/R^2\left(1 - R_g/R\right) = \left(1 + r^2 R_g/R^3\right).$$

In this approximation,

$$ds^2 = \left(1 - \frac{R_g}{\sqrt{z^2 + r^2}}\right)c^2 dt^2 - r^2 d\varphi^2 - \left(1 + \frac{R_g r^2}{(z^2 + r^2)^{3/2}}\right)dr^2 - \left(1 + \frac{R_g z^2}{(z^2 + r^2)^{3/2}}\right)dz^2 - 2\frac{R_g rz}{(z^2 + r^2)^{3/2}}drdz. \quad (2.4)$$

Turning now to the coordinates (2.1) and keeping only linear terms in the radial and vertical deviations from the assumed equilibrium orbit, we get from (2.4):

$$ds^2 = \left(1 - \frac{2gR_0}{c^2}\right)\left(1 + \frac{2g\sin\theta_0}{c^2}x^2 + \frac{2g\cos\theta_0}{c^2}x^3\right)c^2 dt^2 - \left(1 + \frac{2}{r_0}x^2\right)(dx^1)^2 -$$

$$\left[1 + \frac{2gR_0 \sin^2\theta_0}{c^2}\left(1 + \frac{2}{r_0}x^2 - \frac{3}{R_0}(\sin\theta_0 x^2 + \cos\theta_0 x^3)\right)\right](dx^2)^2 -$$

$$\left[1 + \frac{2gR_0 \cos^2\theta_0}{c^2}\left(1 + \frac{2}{z_0}x^3 - \frac{3}{R_0}(\sin\theta_0 x^2 + \cos\theta_0 x^3)\right)\right](dx^3)^2 - \quad (2.5)$$

$$\frac{4gR_0}{c^2}\left[1 + (1 - 3\sin^2\theta_0)\frac{x^2}{r_0} + (1 - 3\cos^2\theta_0)\frac{x^3}{z_0}\right]\sin\theta_0 \cos\theta_0 \cdot dx^2 dx^3.$$

Here we neglect any sine with a power higher than $\sin^2$.

It makes sense to redefine time by introducing $dx^0 = cdt(1 - gR_0/c^2)$. So

$ds^2 = g_{00}(dx^0)^2 + $ other terms of (2.5),

$$g_{00} = 1 + (2g/c^2)(x^2 \sin\theta_0 + x^3 \cos\theta_0). \quad (2.6)$$

$x^0$ is proper time as measured by stationary observers on the Earth's surface. From (2.5) and (2.6), and neglecting deviations $x^2$, $x^3$ from the equilibrium orbit as well as the $g^2$-terms in the final formulas, we have the following vertical components of the Christoffel symbols:

$$\Gamma^3_{00} = \frac{1}{2}g^{33}\left(-\frac{\partial g_{00}}{\partial x^3}\right) = \frac{g\cos\theta_0}{c^2} \tag{2.7}$$

$$\Gamma^3_{11} = \frac{1}{2}g^{32}\left(-\frac{\partial g_{11}}{\partial x^2}\right) = \frac{2g\cos\theta_0}{c^2}, \tag{2.8}$$

where (in our approximation) $g^{32} = -g_{32}(g^{22}/g_{33}) \approx -g_{32}$. Note that $\Gamma^3_{00}$ does not depend on factor $(1-2gR_g/c^2)$ in the original metric component $g_{00}$, and hence on the transformation $c\tau \to x^0$. Note also that $\Gamma^3_{11}$ does not contain any non-gravitational, geometric terms in these cyclindrical coordinates.

The radial components are:

$$\Gamma^2_{00} = \frac{g\sin\theta_0}{c^2}, \tag{2.9}$$

$$\Gamma^2_{11} = -\frac{1}{r_0} + \frac{2g\sin\theta_0}{c^2}. \tag{2.10}$$

$\Gamma^2_{11}$ contains a geometric (centrifugal) term $-1/r_0$. We will first investigate the equilibrium conditions in the horizontal plane, which are the basis of the frozen-spin idea. By using (2.9), (2.10), we can write the condition for the horizontal (radial) equilibrium, $du^2/dc\tau = 0$:

$$\frac{1}{u^1}\frac{du^2}{dc\tau} = \frac{u^1}{r_0} + \frac{e}{mc^2}\left(F^2_1 + F^2_0 u^0/u^1\right) - \frac{g\sin\theta_0}{c^2}\left[\frac{(u^0)^2}{u^1} + 2u^1\right] = 0. \tag{2.11}$$

In the case of $g = 0$, this is the familiar equation $cp = e(B_z - E_r/\beta)r_0$ with $u^0 = \gamma$, $u^1 = \beta\gamma$. (The dynamics of the small oscillations around this equilibrium is beyond the scope of our paper.) The EM fields in (2.11) are solutions of eqs. (1.9), (1.10). For example, neglecting the terms proportional to $\sin^2\theta_0$ they can be:

$$F_1^2 = B\left(1 + x^2/2r_0\right), \quad F_1^3 = Bx^3/2r_0\left(1 + 4gr_0\sin\theta_0/c^2\right), \tag{2.12}$$

$$F_0^2 = E\left(1 - x^2/2r_0\right), \quad F_0^3 = -Ex^3/2r_0\left(1 + 4gr_0\sin\theta_0/c^2\right). \tag{2.13}$$

We assume the longitudinal fields are absent: $F_{23} = 0$, etc. However, we will not use any concrete formulas for $F_k^i$s.

The equations for the labframe radial and vertical spin components, $S^2$, $S^3$, need to deal with the $S^0$-component that appears in the covariant equations. We can use the spin-velocity orthogonality, $S^i u_i = 0$, which leads to $S^0 = -S^1 \dfrac{g_{11} u^1}{g_{00} u^0}$. At the equilibrium kept by some feedback system, $x^3 = 0$, $x^2 = 0$, $g_{00} = 1$, $g_{11} = -1$. From this we get $S^0 u^0 = S^1 u^1$ and, by neglecting the $g^2$ terms,

$$\frac{1}{S^1}\frac{dS^2}{dc\tau} = \frac{u^1}{r_0} + \frac{e}{mc^2}(1+a)\left(F_1^2 + F_0^2 u^1/u^0\right) - \frac{3g\sin\theta_0}{c^2} u^1. \tag{2.14}$$

From (2.11) and (2.14), the frozen-spin condition in the covariant form, $dS^2/dc\tau = 0$, can be written as:

$$\frac{1}{S^1}\frac{dS^2}{d\tau} = \frac{e}{mc}\left[aF_1^2 + \left(a - \frac{1}{(u^1)^2}\right)F_0^2 u^1/u^0\right] + \frac{g\sin\theta_0}{c\beta\gamma} = 0. \tag{2.15}$$

Condition $x^2 = 0$ is included in eq. (2.15), which together with eq. (2.11) defines the connections between ring parameters $r_0$, $u^1$ and fields $F_1^2$, $F_0^2$. $S^1$ in the denominator is permitted because $S^1 \approx s_i \gamma \sim \gamma \neq 0$ in the analyzed experiment. At $g = 0$, (2.15) is the equation for the zero "$g-2$" frequency in a frozen-spin ring (1.1). Unfortunately, eq. (2.15) cannot be satisfied for all particles; the depolarization rate in the *horizontal* plane, $dS^2/d\tau$, can equal zero only on the average over some set of particles, since different particles have different momenta, betatron amplitudes, etc. This means that the beam can be kept polarized only a finite time, let us denote it by $t_p$. Thus, actually, $\langle (dS^2/d\tau)/S^1 \rangle \approx \langle (ds_r/dt)s_l \rangle = 1/t_p \neq 0$, where $s_r$, $s_l$ refer to the radial and longitudinal restframe spin projections. In the case most important for us—the zero vertical magnetic field proposed in [2]—$t_p$ is proportional to the statistical error in keeping the momentum magic:

$$\frac{e}{mc}\left\langle \frac{u^1}{u^0} F_0^2 \left(a - \frac{1}{(u^1)^2}\right)\right\rangle \approx \frac{1}{t_p}, \quad F_1^2 = 0. \tag{2.16}$$

(Here we have neglected the term proportional to $\sin\theta_0$ in (2.15), which is some ten orders of magnitude smaller than any reasonable $1/t_p$.)

We turn now to movement in the vertical direction.

$$\frac{du^3}{dc\tau} + \Gamma_{00}^3 (u^0)^2 + \Gamma_{11}^3 (u^1)^2 = \frac{e}{mc^2}\left(F_0^3 u^0 + F_1^3 u^1\right), \tag{2.17}$$

At the equilibrium $du^3/d\tau = 0$, the left side represents the vertical centrifugal force; the right side represents the EM force balancing the centrifugal one. In the cylindrical coordinates, the latter is proportional to $g$:

$$\Gamma_{00}^3 (u^0)^2 + \Gamma_{11}^3 (u^1)^2 = (g/c^2)\cos\theta_0 \left((u^0)^2 + 2(u^1)^2\right) = \gamma^2 (g/c^2)\cos\theta_0 (1 + 2\beta^2). \tag{2.18}$$

Condition $x^3 = 0$ imposes a connection between $u^1$ and the constant components of the fields $F_0^3$, $F_1^3$ i.e., the components independent of $x^3$. Indeed, the EM fields on the right side of (2.17) obviously do not equal zero at the vertical equilibrium, $du^3/dc\tau = u^3 = 0$, since they balance the vertical gravitational acceleration given on the left side. By design, these EM fields contain some field gradients. For example, $F_0^3 \propto \left(E_{z0} + (\partial E_z/\partial x^3)_0 x^3\right)$.

To keep $x^3 = 0$, that is, to keep the beam at the center of the vacuum chamber, requires a non-zero constant component $E_{z0}$. It can be provided by design. This field must of course be taken into account in the vertical spin equation.

In contrast to the horizontal spin component (which is frozen, by design), the vertical spin component, $S^3$, is not constant in time:

$$\frac{dS^3}{dc\tau} + \left(\Gamma^3_{00} + \Gamma^3_{11}\right)S^1 u^1 = \frac{e}{mc^2}(1+a)\left[F^3_0 \frac{u^1}{u^0} + F^3_1\right] \cdot S^1, \qquad (2.19)$$

where we take into account that $S^0 u^0 \approx S^1 u^1$.

The *direct* spin-gravitation interaction is described by the left side of (2.19). The right side describes the *indirect* spin-gravity interaction through the EM field balancing the force of gravity. If we multiply (2.19) by $u^1/S^1$ and subtract the result from (2.17), we get at $du^3/d\tau = 0$:

$$\frac{u^1}{S^1}\frac{dS^3}{dc\tau} + \Gamma^3_{00}\left((u^1)^2 - (u^0)^2\right) = \frac{e}{mc^2} u^1 \left\{aF^3_1 + \frac{u^1}{u^0} F^3_0 \left[a - \frac{1}{(u^1)^2}\right]\right\}. \qquad (2.20)$$

(Being written in a general form, eqs. (2.17), (2.19) and (2.20) do not depend on the choice of coordinate system.) We see from (2.20) that in a frozen-spin ring, if the momentum is magic, $1/(u^1)^2 = a$ and $F^2_1 = 0$, and if the radial magnetic field is absent, $F^3_1 = 0$, then the right side of (2.20) equals zero *independently of the electric field*. The left side of (2.20) depends only on $\Gamma^3_{00}$, not $\Gamma^3_{11}$. We use this result below and in Sections 3 and 4.

With magic momentum and no magnetic field, it follows from (2.7), (2.16) and 2.20) that the rotation rate of the restframe spin, $s_l \approx S^1/\gamma$, $s_z \approx S^3$, as a function of the labframe time, $t \approx \gamma\tau$, is:

$$(ds_z/dt)_{grav} \approx (g\cos\theta_0/c)\sqrt{a} \cdot s_l + \frac{1}{t_p}\frac{E_z}{E_r}s_l. \tag{2.21}$$

From (2.17) and (2.18), $E_z/E_r \sim gr_0/c^2 \ll 1$; in addition, $r_0/ct_p \ll 1$. The last term of eq. (2.21) can therefore be neglected in the cylindrical coordinates. Also, $\cos\theta_0 \approx 1$, so

$$(ds_z/dt)_{grav} = (g/c)\sqrt{a}\,s_l. \tag{2.22}$$

The spin-gravity and EDM rotations are observed in the same ring:

$$(ds_z/dt)_{edm} = \eta(eE_r/2mc) \cdot s_l. \tag{2.23}$$

## 3. Spin rotation: spherical coordinates

In a system of spherical coordinates, the "vertical" is now directed along the radius vector from the Earth's center to some point of the orbit. It makes sense, again, to remove the constant $2\Phi_0/c^2$ from $(1 - R_g/R)$ in eq. (1.7) by redefining time (as in the cylindrical coordinates case), $x^0 = ct\sqrt{1+2\Phi_0/c^2} \cong ct(1-gR_0/c^2)$, and to neglect the second-order term $(\Phi_0/c^2)(mg\bar{x}^3/c^2)$ in the transformed $g_{00}$. (As noted in Section 2, such a transformation does not influence $\Gamma^3_{00}$.) Then we introduce the coordinates in the vicinity of the ring:

$$x^0, \ x^1 = (R_0\sin\theta_0)\varphi \equiv r_0\varphi, \ \bar{x}^2 = R_0(\theta - \theta_0), \ \bar{x}^3 = R - R_0. \tag{3.1}$$

(See $\vartheta = \theta - \theta_0$ in Fig. 1.) This gives us:

$$ds^2 = \left(1 + \frac{2g\bar{x}^3}{c^2}\right)(dx^0)^2 - \left(1 + \frac{2\bar{x}^3}{R_0} + \frac{2\bar{x}^2 \cos\theta_0}{r_0}\right)(dx^1)^2 -$$
$$\left(1 + \frac{2\bar{x}^3}{R_0}\right)(d\bar{x}^2)^2 - \left(1 + \frac{2gR_0}{c^2} - \frac{2g\bar{x}^3}{c^2}\right)(d\bar{x}^3)^2.$$

(3.2)

The Christoffel symbols are:

$$\Gamma^3_{00} = \frac{g}{c^2}; \quad \Gamma^3_{11} = -\frac{\sin\theta_0}{r_0} + \frac{2g}{c^2},$$

(3.3)

$$\Gamma^2_{00} = 0; \quad \Gamma^2_{11} = -\frac{\cos\theta_0}{r_0}.$$

(3.4)

A *vertical* non-gravitational centrifugal force equal to $(u^1)^2/R_0$ has now appeared in eq. (2.17)—as a part of the term $\Gamma^3_{11}(u^1)^2$. Acting along the new vertical direction, it corresponds to the term $-\sin\theta_0/r_0 = -1/R_0$ in $\Gamma^3_{11}$ of eq. (3.3). That geometric term—which is absent in (2.8) of the cylindrical coordinates and is larger than the purely gravitational $g/c^2$ by a factor of $\frac{c^2}{gR_0} \approx 10^9$ —is simply the projection of the usual cylindrical horizontal centrifugal force onto the new vertical axis. The above factor is even larger in comparison with the EDM spin rotation. The appearance of $\cos\theta_0$ in the $\Gamma^2_{11} = -\cos\theta_0/r_0$ of eq. (3.4) results from projecting that same force onto the new horizontal axis. It follows from eq. (3.3) that an extremely small change of direction of the vertical axis, $\Delta\theta = \frac{gr_0}{c^2} \sim 10^{-15}$ rad, will cause a geometric spin rotation that imitates the EDM and can exceed both the gravitational and the assumed EDM rotations.

The sensitivity of observed spin rotation to one's choice of vertical direction would be no experimental problem if we could control the vertical with an accuracy of $10^{-15}$ rad. However, such accuracy is currently impossible. This can lead to considerable uncertainty about systematic errors in the frozen-spin EDM experiment and about the source of the spin rotation in general. From (2.20), which is correct in any coordinate system, follow the conditions that solve this problem: *magic momentum and no magnetic field*:

$$\vec{B} = 0 \quad \text{and} \quad \beta\gamma = 1/\sqrt{a} . \tag{3.5}$$

From (3.5):

$$\frac{1}{S^1}\frac{dS^3}{d\tau} = \Gamma^3_{00} c/u^1 + \frac{e}{mc} F^3_0 \frac{u^1}{u^0}\left(a - \frac{1}{(u^1)^2}\right) = \\ \Gamma^3_{00} c/u^1 + \left[\frac{e}{mc} F^2_0 \frac{u^1}{u^0}\left(a - \frac{1}{(u^1)^2}\right)\right]\left(\frac{F^3_0}{F^2_0}\right) \tag{3.6}$$

The frozen-spin condition in the absence of any magnetic field takes the form:

$$\frac{1}{S^1}\frac{dS^2}{d\tau} \approx \frac{e}{mc} F^2_0 \frac{u^1}{u^0}\left(a - \frac{1}{(u^1)^2}\right) \approx 0 , \tag{3.7}$$

where we neglect terms proportional to $g/c$. (Any such term in the equation for the horizontal spin component is much smaller than the statistical and systematic errors caused by other, non-gravitational terms which we also do not take into account.) Thus,

$$\frac{1}{S^1}\frac{dS^3}{d\tau} = \Gamma^3_{00} c/u^1 + \left(\frac{dS^2}{S^1 d\tau}\right)\left(\frac{F^3_0}{F^2_0}\right). \tag{3.8}$$

In the approximation $\cos\theta_0 = 1$, the right side of (3.8) equals the right side of (2.21), since $\Gamma^3_{00} = g/c^2$ and $(dS^2/S^1 d\tau)(F^3_0/F^2_0) \approx (1/t_p)(E_z/E_r)$ in (3.8). The product of two factors, $(1/t_p)$ and $(E_z/E_r)$, can be made smaller than the main term, $\Gamma^3_{00} c/u^1$

$= (g/c)\sqrt{a}$. Indeed, from the equations of the vertical and horizontal motions,

$E_z / E_r \sim \Gamma^3_{11} / \Gamma^2_{11} \sim r_0 / R_0 = \sin\theta_0$. Therefore, the condition

$$(1/t_p)(E_z / E_r) \ll (g/c)\sqrt{a} \qquad (3.9)$$

becomes in spherical coordinates,

$$(1/t_p)\sin\theta_0 \ll (g/c)\sqrt{a}. \qquad (3.10)$$

(3.10) is rather easy to satisfy in a purely electric ring. If (3.9) is satisfied, then the formula (2.22) for spin-gravity rotation in a purely electric frozen-spin ring is correct in both cylindrical and spherical coordinate systems.

### 4. Spin rotation: isotropic-spherical coordinates

We now need to investigate whether spin rotation is sensitive to small variations of the chosen vertical coordinates when the vertical direction is given. The isotropic-spherical coordinate system seems to be a good tool for such an investigation. In the linear approximations with respect to $g$, the isotropic-spherical coordinates are $\varphi$, $\theta$, $\rho$, where the vertical scale

$$\rho = R - R_g / 2 \qquad (4.1)$$

is different from the vertical scale, $R$, of the spherical coordinate system. So the Schwarzschild metric becomes:

$$ds^2 = (1 - R_g / \rho)c^2 dt^2 - (1 + R_g / \rho)\left[\rho^2\left(\sin^2\theta d\varphi^2 + d\theta^2\right) + d\rho^2\right]. \qquad (4.2)$$

At the equilibrium $\theta = \theta_{eq} = \theta_0 = Arc\sin(r_0 / R_0)$, that is, our "vertical direction" here is the same as that of the usual spherical coordinate system, Section 3. Further, $\rho = \rho_{eq} = R_0 - R_g/2$, $R_g \approx 2g\rho_0^2/c^2$. From (4.2), neglecting terms proportional to small deviations from the equilibrium,

$$\Gamma^\rho_{tt} = g/c^2 \tag{4.3}$$

$$\Gamma^\rho_{\varphi\varphi} = \left(-1/\rho_{eq} + g/c^2\right)\sin^2\theta_0. \tag{4.4}$$

This is similar to the Christoffel symbols in the spherical coordinates, eqs. (3.3), but without factor 2 before $g/c^2$, which is present in the spherical Christoffel component. The Christoffel symbols for horizontal movements in the spherical coordinates remain the same for those movements in the isotropic coordinates (taking into account notational differences). Eq. (4.4) without factor 2 and with substitutions $\Gamma^\rho_{tt} \to \Gamma^3_{00}$, $\Gamma^\rho_{\varphi\varphi} \to \Gamma^3_{11}$ can also be obtained, by using the small deviations from the equilibrium in the isotropic-spherical coordinates:

$$x^1 = \rho_{eq} \cdot \sin\theta_0 \cdot \varphi, \quad x^2 = \rho_{eq}(\theta - \theta_0), \quad x^3 = (\rho - \rho_{eq}). \tag{4.5}$$

Here we will omit the corresponding calculations.

In a ring satisfying conditions (3.5)—that is, a purely electric frozen-spin ring—the difference between the $\Gamma^3_{11}$ of the spherical and isotropic-spherical coordinate systems cannot lead to any uncertainty about observed spin rotations. First, the contribution of $\Gamma^3_{11}$ to spin rotation is cancelled in the isotropic-spherical system in the same way as in the other two coordinate systems, eq. (2.20). In the linear approximations with respect to $g$, $\Gamma^3_{00}$ and the products $\Gamma^3_{00}(u^0)^2 \propto g\gamma^2$, $\Gamma^3_{00}(u^1)^2 \propto g(\beta\gamma)^2$ are the same for all three coordinate systems considered here. Therefore, the spin and momentum accelerations caused by $\Gamma^3_{00}$ are also the same for all systems. Second, although the isotropic-spherical $\Gamma^3_{11}$ differs from the usual spherical $\Gamma^3_{11}$, the vertical accelerations, $\Gamma^3_{11} v^2$, are the same due to the different expressions for particle velocity, $v$, in these two coordinate systems. In the isotropic-spherical system, $v^2 = \rho^2_{eq} \dot{\varphi}^2 \sin^2 \theta_0$; in the spherical, $v^2 = R_0^2 \dot{\varphi}^2 \sin^2 \theta_0$. From this,

$$(\Gamma^3_{11} v^2)_{isotr} = \left(-\frac{1}{\rho_{eq}} + \frac{g}{c^2}\right) \rho^2_{eq} \dot{\varphi}^2 \sin^2 \theta_0 = \left(-(R_0 - R_g/2) + \frac{g \rho^2_{eq}}{c^2}\right) \dot{\varphi}^2 \sin^2 \theta_0 = \left(-R_0 + 2\frac{g R_0^2}{c^2}\right) \dot{\varphi}^2 \sin^2 \theta_0 = \left(-\frac{1}{R_0} + 2\frac{g}{c^2}\right) R_0^2 \dot{\varphi}^2 \sin^2 \theta_0 = (\Gamma^3_{11} v^2)_{sph}. \quad (4.6)$$

Thus, in our approximation, the dynamics of the vertical spin components and the vertical momentum components are the same in the spherical and isotropic-spherical systems of coordinates.

**5. Conclusions**

Our main result is that spin-gravity rotation in the vertical tangential plane can be estimated by eq. (2.22) under conditions (3.5), (3.9) and (3.10). Since (2.22) is designed to be applied to coordinate systems with different definitions of the vertical, its relative accuracy is only $(r_0 / R_0)^2$. At this level of accuracy one could get the same result (2.22) by assuming a uniform gravitational field all along the ring ($\cos \theta_0 = 1$) and then using special relativity in an accelerated frame plus the equivalence principle (a uniform gravitational field is equivalent to an accelerated frame). However, such an approach would limit the generality of the method without simplifying the calculations. For this reason we have used Riemannian geometry in the frame of general relativity. The result, in addition to (2.22), is a set of equations (2.7)-(2.10), (3.3), (3.4) containing higher-order terms that include the effects of geodetic precession [11]. These terms are the basis for corrections to eq. (2.22) of the order $(r_0/R_0)^2$. We do not analyze them because they are beyond the accuracy level of current measurements.

As noted in Section 1, we do not take into account the Earth's rotations; in particular, we neglect the gravitational field generated by the angular momentum of the Earth, which gives rise to the Lense-Thirring precession (the dragging of inertial frames) [11]. The relative contribution of the Lense-Thirring precession to our (2.22) would be of the order $v/c$, where $v$ is the spin velocity of the Earth's surface. Experimentally, this effect is well beyond the accuracy level of the proposed EDM measurements.

Our calculations of spin rotation in different coordinate systems reveal that a big nongravitational, geometric effect—a *vertical* centrifugal force—is a source of major ambiguity concerning observed rotations in the vertical plane tangential to the orbit. Depending on how vertical is defined by the experimental setup one uses, these spin rotations can range from zero (the vertical direction associated with cylindrical coordinates) to some nine orders of magnitude bigger (the vertical direction associated with spherical Schwartzschild coordinates) than the purely gravitational spin rotation, eq. (2.22). In order to assess a source of systematic error in EDM measurement, one must control the experimental choice of vertical direction with accuracy $10^{-15}$ rad. That is currently difficult. This means that one cannot interpret the observed spin-rotation values; one is, in effect, at the mercy of the ambiguity described above.

Following from our eq. (2.20), the solution to this dilemma is to bypass the ambiguity by reducing its influence. Indeed, our calculations show that the influence of the definition of vertical on observed spin rotations can be virtually cancelled if one uses a purely electric frozen-spin ring, which is possible for particles with positive anomalous gyromagnetic factors, $a > 0$. (It is impossible for deuterons, with their $a = -0.14$.) As we have demonstrated, under condition (3.9) in a purely electric frozen-spin ring, the rate (2.22) of gravitational spin rotations in the vertical plane can be the same in different coordinate systems, despite a big difference in the experimentally chosen directions of the vertical axes, $\Delta\theta \sim \theta_0 \sim r_0 / R_0$.

The physics behind this canceling of geometric spin rotations is the following. The vertical centrifugal force is canceled in the eqs. of motion by the EM fields balancing that force, eq. (2.17). However, it retains its hold on the vertical spin component due to those very electric and magnetic fields, eq. (2.20). But if $p$ is magic, then the influence of the (vertical) electric field on the vertical component of spin is blocked by the factor $\left[a - (mc)^2 / p_{mag}^2\right] = 0$, since the particle momentum is magic in the frozen-spin electric ring. Also, the choice of $p = p_{mag}$ requires that $B_z = 0$. The additional condition $B_r = 0$ pulls the radial magnetic field, $B_r$, out of eq. (2.20). In the absence of any magnetic field, only the vertical electric field balances the vertical centrifugal force; but, as explained above, this fact does not influence spin.

Applied to the proposed proton EDM experiment [2], these results strongly suggest that magnetic fields (magnetic focusing, for example) should be avoided in the frozen-spin EDM ring. The spin rotation by gravity, eq. (2.22), will still be bigger than the expected sensitivity to the $10^{-29} e \cdot cm$ EDM. However, interpretive ambiguity will be avoided. We will instead have a well-defined systematic error that can be simply extracted numerically from the observed spin-rotation rate by using our formula (2.22).

For protons, $a = 1.79$, this formula gives $\left(ds_z / dt\right)_{grav} = s_l \times 43.8$ nrad/s. Keeping $\vec{B} = 0$ without serious constraints on $\vec{E}$ is still a challenge, but much easier than (say) keeping $B_r$, $E_z$, $B_z$ and $E_r$ at some precise non-zero values, as a non-electric ring would require.

Applied to investigation of the gravity-spin effect itself, our results show that the spin-gravity effect can be unambiguously measured and that a purely electric frozen-spin ring is a way to do so. Electrons would be a good choice of test particle in a ring exclusively dedicated to measuring the spin-gravity effect, because the magic momentum $p_{mag}$ of electrons is small, which means that the ring can be small. According to (2.22), the theoretical expectation is that this rate for electrons, $a = 0.00116$, will be $\left(ds_z / dt\right)_{grav} = s_l \times 1.1 \times 10^{-9}\, s^{-1}$. A systematic error caused by the (still unknown) electron EDM will probably not be larger than these spin-gravity rotations. As for the radiation processes, the effects of radiative polarization [12] and depolarization [13] for electrons with energy 15 MeV in a purely electric frozen-spin ring are very small and can be neglected.

Apart from its intrinsic interest, the spin-gravity experiment proposed here might constrain modified gravitational theories for which the equivalence principle fails to hold, as well as theories involving torsion in which elementary particle spin does not couple to gravity in the same way as it does in general relativity, i.e., as macroscopic angular momentum.

**Acknowledgments**

The work of one of the authors (E.F.) was supported by NSF Grants PHY-0757735 and PHY-1068541, and by the hospitality of the Theoretical Astrophysics Including Relativity Group at Caltech, and the Department of Applied Mathematics and Theoretical Physics at the University of Cambridge.

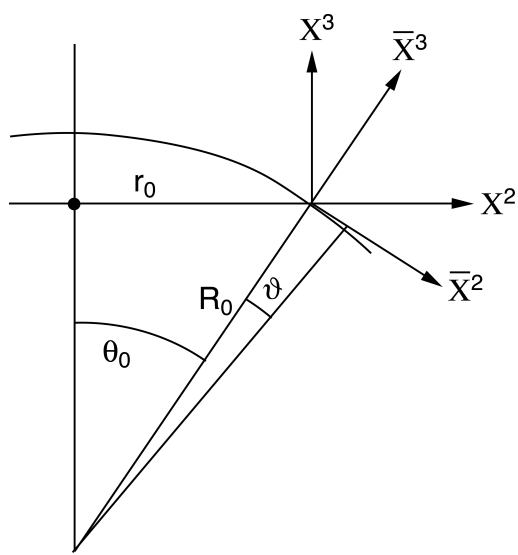

**Fig. 1**. General scheme of the EDM ring. The ring lies at the Earth's spherical surface. The Earth's radius is $R_0$; the ring radius is $r_0$. The cylindrical contravariant coordinates of a particle are denoted by $x^1$ (along the orbit, not visible here), $x^2$ (along the instant radius of the orbit at point $x^1$), and $x^3$ (orthogonal to both). The corresponding spherical coordinates are $\vec{x}^1 = x^1, \vec{x}^2, \vec{x}^3$.